\documentclass[aps,prb,twocolumn,amsmath,amssymb,superscriptaddress,nofootinbib]{revtex4}
\usepackage{epsfig}

\begin{document}



\title{Classical Spin Models with Broken Continuous Symmetry: Random Field Induced Order and Persistence of Spontaneous Magnetization}


\author{Aditi Sen(De)}
\affiliation{ICFO-Institut de Ci\`encies Fot\`oniques,
Parc Mediterrani de la Tecnologia,
E-08860 Castelldefels (Barcelona), Spain}
\author{Ujjwal Sen}
\affiliation{ICFO-Institut de Ci\`encies Fot\`oniques,
Parc Mediterrani de la Tecnologia,
E-08860 Castelldefels (Barcelona), Spain}
\author{Jan Wehr}
\affiliation{ICFO-Institut de Ci\`encies Fot\`oniques, Parc
Mediterrani de la Tecnologia, E-08860 Castelldefels (Barcelona),
Spain} \affiliation{On leave from Department of Mathematics,
University of Arizona, Tucson, AZ 85721-0089, USA}
\author{Maciej Lewenstein}
\affiliation{ICREA-Instituci{\'o} Catalana de Recerca i Estudis Avan{\c c}ats, Lluis Companys 23, 08010 Barcelona, Spain}
\affiliation{ICFO-Institut de Ci\`encies Fot\`oniques,
Parc Mediterrani de la Tecnologia,
E-08860 Castelldefels (Barcelona), Spain}


\begin{abstract}

We consider a classical spin model, of two-dimensional spins,
 with continuous symmetry, and investigate the effect of a symmetry breaking
unidirectional quenched
disorder on the magnetization of the system. We work in the mean field regime. We show, by numerical simulations and by
perturbative calculations in the low as well as in the high temperature limits, that although the continuous symmetry of
the magnetization is lost, the system still magnetizes,
albeit with a lower value as compared to the
case without disorder. The critical temperature at which the system starts magnetizing, also decreases with the
introduction of disorder. However, with the introduction of an additional constant magnetic field, the component of
magnetization in the direction  that
is transverse to the disorder field increases with the introduction of the quenched disorder.
We discuss the same effects also for three-dimensional spins.
\end{abstract}


\maketitle

\section{Introduction}
\label{sec:introduction}

Disordered systems, both classical and quantum,  lie at the center-stage of condensed matter physics  
\cite{booksondisorder, amrasobai}.
Challenging 
open questions in disordered systems include those in the realms of  spin glasses \cite{spinglass}, neural networks \cite{nn},
percolation \cite{percolation}, and high $T_c$ superconductivity \cite{superconduc}. 
Phenomena like Anderson localization \cite{Anderson,booksonanderson}, and absence of magnetization in several 
classical spin models \cite{imryma1, imryma2,janek} are effects of disorder.

In particular, classical ferromagnetic spin models with discrete, or continuous, symmetries are very sensitive to random magnetic 
fields, distributed 
in accordance with the symmetry, in low dimensions \cite{imryma1}. For instance, an
arbitrary small random magnetic field with \(Z_2\) (\(\pm\)) symmetry  
destroys spontaneous magnetization in the Ising model in 2D at any temperature \(T\), including \(T=0\). 
Similar effect holds for the XY model in 2D at \(T=0\) in a random field 
with \(U(1)\) (\(SO(2))\) symmetry, or Heisenberg model in 2D at \(T=0\) in \(SU(2)\) (\(SO(3)\)) symmetry in random field. In these cases, the effects of disorder amplify the effects of continuous symmetry, that destroys spontaneous magnetization at any \(T>0\). The effect is even more 
dramatic in 3D, where the random field destroys spontaneous magnetization at any \(T \geq 0\). (For a general description of these, see
\cite{imryma1, imryma2, janek}.)

The appropriate symmetry of the random field is essential for the above mentioned results. The natural question arises as to what happens if 
the distribution of the random field does not exhibit the symmetry, in particular the continuous symmetry. Yet another natural question is 
how does the spin systems in random fields behave in the quantum limit. The latter question is particularly interesting in view of the fact 
that nowadays it is possible to realize practically ideal models of quantum spin systems (with spin s=1/2, 1, 3/2, ..., and 
with Ising, XY, or Heisenberg interactions) in controlled random fields \cite{amrasobai, Armand}. 
It is therefore very important to understand the physics of  both classical and quantum spin models in random fields that break their
symmetry.

In this paper, we will consider the classical XY spin model in a random field that breaks the continuous U(1) (\(SO(2)\)) symmetry. 
We investigate this model in the
mean field approximation \cite{mean}.  Despite its simplicity, this model with
the two-dimensional spin variable, magnetizes in the absence of disorder 
below a certain critical temperature, which can be calculated exactly.  As a
result of continuous symmetry, the possible values of spontaneous
magnetization form a circle in the plane.  A random magnetic field
pointing along the \(Y\)-direction is introduced, by adding a new term
to the energy of the model.  This term breaks the continuous
symmetry of the model, but the critical temperature persists.  The
present paper studies the critical behaviour and properties of
spontaneous magnetization in the resulting mean-field disordered
system.  We prove that, as may be expected, adding a random field
lowers the critical temperature.  Next, we show that the
magnetization of the disordered system is lower than that of the
pure one---another intuitively plausible result, first shown
numerically and then by perturbation expansions: one around the
critical temperature of the pure model and another for low
temperatures.

Next, we introduce a \emph{constant} magnetic field, which breaks
the continuous symmetry of the model even in the absence of
disorder.  In fact, the system now magnetizes at all temperatures
in the direction parallel to the magnetic field.  When we now also
add a random field as described above, the length of the
magnetization vector decreases again. Moreover, the magnetization
gets atrracted towards the \(X\)- axis, i.e. the direction
transverse to that of the random field. However, the
\(X\)-component of the magnetization \textbf{can} increase for
certain choices of the constant field.  We view this effect as a
case of ``random field induced order'', by analogy with the effect
studied in \cite{Armand}, where numerical evidence was given for
appearance of magnetization in the XY model on a two-dimensional
lattice with the introduction of disorder.  In contrast to the
present work, in this other case no mean-field approximation was
used and no uniform magnetic field was introduced.

The effect of ``random field induced order'' has, of course, a long history \cite{history}. Recently it has become vividly discussed 
in  the context of $XY$ ordering in a graphene quantum Hall ferromagnet \cite{lee}, and ordering in 
$\phantom{i}^3$He-A arerogel and  amorphous ferromagnets \cite{fomin}. Let us stress that the novel aspects of our previous  paper \cite{Armand} consists in clarifying certain aspect of the rigorous proof of the appearance of magnetization in XY model at $ T=0 $, presentation of a novel evidence for the same effect at $ T> 0$, and a proposal for realisation of quantum version of the effect with ultracold atoms. In  the subsequent paper \cite{Armand2}, 
we have shown how the ``random field induced order'' exhibits itself in a system of two-component trapped Bose-Einstein condensate with random Raman inter-component coupling. The novelty of the present paper lies in systematic mean field treatment of the disordered $XY$ model 
with particular emphasis on the response to the constant magnetic field.  

The paper is arranged as follows. In Sect. \ref{sec-bapi_bari_ja}
the ferromagnetic XY model is introduced.  A symmetry breaking
random field is added in Sect. \ref{Delhi_abhi_door_hai} and the
results of numerical simulations of and perturbative calculations
on the resulting model are presented. In Sect. \ref{girgiti}, 
study the system with an additional constant field and, in
particular, show presence of random field-induced order.  We
discuss the results in Sect. \ref{alochona}, arguing in particular
that the analogs of our results will hold for the mean-field
version of the classical Heisenberg model.

\section{Ferromagnetic XY model: Mean Field approach}
\label{sec-bapi_bari_ja}

Consider a lattice, each site \(i\) of  which is occupied by a
``spin'', which is a unit vector $\vec{\sigma}_{i} = (\cos
\theta_i, \sin \theta_i)$ on a two-dimensional plane (called the
XY plane). The nearest-neighbor ferromagnetic XY model is defined
by the Hamiltonian
\begin{equation}
\label{eq_hamil}
H_{XY} = - J \sum_{|i - j| = 1} {\vec{\sigma_i}} \cdot \vec{\sigma_j},
\end{equation}
with a coupling constant $J > 0$.
This model does not have any spontaneous magnetization, at any
temperature, in one and two dimensions (Mermin-Wagner-Hohenberg
theorem \cite{Mermin-Wagner}), while a finite magnetization
appears in higher dimensions for sufficiently low temperatures
\cite{spinwave, Frohlich}.

Let us assume that the total number of spins in our system is $N$.
In the mean field approximation
every spin is assumed to interact with all other spins (not just
with the nearest neighbors) with the same coupling constant $-J$.
Therefore, the contribution of the spin at $i$ to the total energy
of the system equals
\[
\left(-{J \over N}\sum_{j:j\ne i} \vec{\sigma}_j \right) \cdot
\vec{\sigma}_i,
\]
where we divided the energy term by $N$ in order to preserve its
order of magnitude. This effective interaction, replacing the
nearest neighbor interaction in \(H_{XY}\), is for large $N$
approximately equal
\begin{eqnarray}
 &&\frac{1}{N}\left( -J \sum_{j} {\vec{\sigma_j}} \right) \cdot {\vec{\sigma_i}} \nonumber \\
 = && - J {\vec{m}} \cdot {\vec{\sigma_i}},
\end{eqnarray}
where ${\vec{m}} = \frac{1}{N}\sum_{i=1}^{N} \vec{\sigma}_i$. The
mean field approximation consists of treating $\vec{m}$ as a
genuine constant vector and adjusting it so, that the canonical
average of the spin at (any) site $i$ equals this constant.
If the system is in canonical equilibrium at temperature \(T\),
the average value of the spin vector \(\vec{\sigma}_i\) is
\begin{equation}
\langle {\vec{\sigma_i}}\rangle =
\frac{\int {\vec{\sigma}_i} \exp(\beta J {\vec{m}}\cdot {\vec{\sigma}_i}) d{\vec{\sigma}_i}}
                                 {\int \exp(\beta J {\vec{m}}\cdot {\vec{\sigma}_i}) d{\vec{\sigma}_i}},
\label{sigma_mag}
\end{equation}
where $\beta = 1/(k_B T)$, with $k_B$ being the Boltzmann constant.
This average is independent of the site $i$. Consistency requires
that the left hand side (l.h.s.) of the above equation be equal to
the magnetization ${\vec{m}}$.  Hence, we obtain the mean field
equation
\begin{equation}
 {\vec{m}} = \frac{\int {\vec{\sigma}} \exp(\beta J {\vec{m}}\cdot {\vec{\sigma}}) d{\vec{\sigma}}}
                                 {\int \exp(\beta J {\vec{m}}\cdot {\vec{\sigma}}) d{\vec{\sigma}}},
\label{meanfield_mag}
\end{equation}
where we have dropped the index \(i\).  Equations of this type,
for various modifications of the original interaction are the main
subject of this work.

Let
\begin{equation}{\vec{m}} = (m \cos a, m \sin a).\end{equation}
For sufficiently high temperatures, the only solution fo the mean
field equation is $\vec{m} = 0$.  There exists a $\beta_c^0$, such
that for $\beta > \beta_c^0$, this system magnetizes. (Later, we
will consider the case of a system with an additional quenched
disordered field of strength \(\epsilon\). The superscript of
\(\beta_c^0\) is anticipation of that case.) By symmetry, the
solutions of the above mean field equation (Eq.
(\ref{meanfield_mag})) form a circle
\begin{equation}
 \label{dhorimaachh}
|{\vec{m}}| = m_0
,
\end{equation}
with a strictly positive radius $m_0$, for any $\beta >
\beta_c^0$. Choosing the phase \(a=0\), Eq. (\ref{meanfield_mag})
reduces to
\begin{equation}
 m = \frac{\int_{0}^{2 \pi} \cos\theta \exp(\beta J m \cos\theta) d\theta}
                                 {\int_{0}^{2 \pi} \exp(\beta J m \cos\theta) d\theta},
\label{meanfield_mag_simplify}
\end{equation}
where we have taken
\begin{equation}\vec{\sigma} = (\cos\theta, \sin\theta). \end{equation}
 \begin{figure}[tbp]
\begin{center}
\epsfig{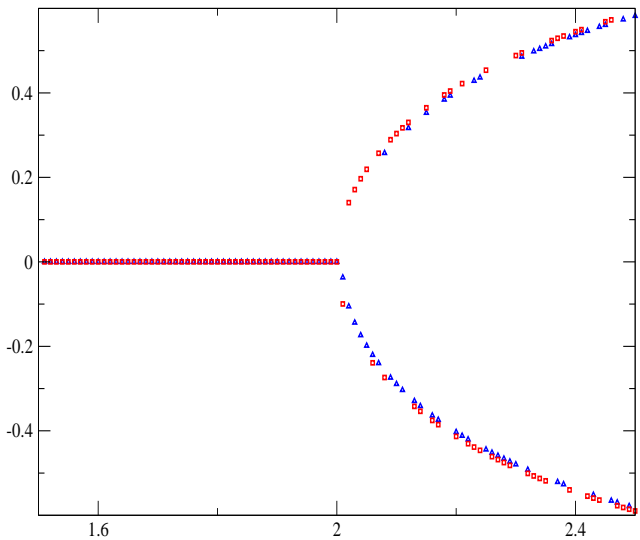}
\caption{(Color online.) The behavior of magnetization (vertical axis) with respect to $\beta J$ (horizontal axis).
The (red) squares represent the magnetization of the XY model without disorder (the corresponding Hamiltonian being \(H_{XY}\)),
while
the (blue) traingles are that for the same, but with disorder (the corresponding Hamiltonian being \(H_{\epsilon}\) with
\(\epsilon =0.1\)).}
\label{comparison_hightemp}
 \end{center}
\end{figure}
In Fig. \ref{comparison_hightemp}, the red squares represent the
cross-section, of the surface of solutions of Eq.
(\ref{meanfield_mag}) in the \((m \cos a, m \sin a, \beta)\)
Cartesian space, in the $\cos a = 0$ plane.


From numerical simulations (see Fig. \ref{comparison_hightemp}), we found that $\beta_c^0 J\approx 2.00$.
One can show analytically that \(\beta_c^0 J\) is exactly 2, as follows.
Let us denote the right hand side (r.h.s.) of Eq.
(\ref{meanfield_mag_simplify}) by $F(m)$. The condition for the
system to magnetize, for a given value of $\beta$, is that the
derivative of the r.h.s. of Eq. (\ref{meanfield_mag_simplify}) at
$m = 0$, i.e. $F^{'}(m){|}_{m = 0}$ should be greater than the
derivative of the l.h.s.
It is easy to check that
$$F^{'}(m){|}_{m = 0} = \beta J/2.$$ This implies that the system
possesses a nonzero magnetization, if and only when $$\beta > 2/J
= \beta_c^0.$$

\section{Ferromagnetic XY model in a random field}
\label{Delhi_abhi_door_hai}

We will now consider the effect of additional quenched random
fields.  Let us begin by reminding the notions of quenched
disorder and quenched averaging.

\subsection{Quenched averaging}

The disorder considered in this paper is ``quenched'', i.e. its
configuration remains unchanged for a time that is much larger
than the duration of the dynamics considered. In the systems that
we study, it is the local magnetic fields that are disordered.
They are random variables following certain probability
distributions. Since the disorder is quenched, a particular
realization of all the random variables remains fixed for the
whole time necessary for the system to equilibrate.  An average of
a physical quantity, say \(A\), is thus to be carried out in the
following order.
\begin{itemize}

\item[(a)] Compute the value of the physical quantity \(A\), with
the fixed configuration of the disorder.

\item[(b)] Average over the disordered parameters.

\end{itemize}

This mode of averaging is called ``quenched'' averaging. It may be
mentioned that an averaging in which items (a) and (b) are
interchanged in order, is called ``annealed'' averaging.
Physically it corresponds to a situation when the disorder
fluctuates on time scales comparable to the system's thermal
fluctuations.

\subsection{The model and the mean field equation for magnetization}

The $XY$ model with an inhomogeneous magnetic field has the
interaction:
\begin{equation}
\label{XY with magnetic field} H = - J \sum_{|i-j| =
1}\vec{\sigma_i} \cdot \vec{\sigma_j} - \epsilon \sum_i\vec{h}_i
\cdot \vec{\sigma}_i.
\end{equation}
where the two-dimensional vectors $\vec{h}_i$ are the external
magnetic fields, up to a coefficient $\epsilon$.  In the sequel
$\vec{h}_i$ are random variables of order one, they model the
disorder in the system and thus $\epsilon$ measures the disorder's
strength.  More precisely, let $\vec{h}_i$ be independent and
identically distributed random variables (vector-valued).  We want
to study the effect of including such a random field term in the
$XY$ hamiltonian at small values of $\epsilon$.  As argued in
\cite{Armand}, in lattice $XY$ models this effect depends
critically on the properties of the probability distribution of
the random fields.

If the distribution of the $\vec{h}_i$ is invariant under
rotations, there is no spontaneous magnetization at any nonzero
temperature in any dimension $d \leq 4$.  \cite{imryma1, imryma2, janek}

We now want to see the effect of a random field that does not have
the  rotational symmetry of the XY model interaction (\ref{eq_hamil}), considering the case when
\begin{equation}
\vec{h}_i = \eta_i \cdot \vec{e}_y
\end{equation}
where $\eta_i$ are scalar random variables with a distribution
symmetric about $0$ and $\vec{e}_y$ denotes the unit vector in the
$y$ direction.  The main result of \cite{Armand} is that on the
two-dimensional lattice such a random field will break the
continuous symmetry and the system will magnetize, even in two
dimensions, thus destroying the Mermin-Wagner-Hohenberg effect.
Above two dimensions the pure $XY$ model magnetizes at low
temperatures and it has been suggested in \cite{Armand} that the
uniaxial random field as described above may enhance this
magnetization.  In the present paper we want to study related
effects at the level of a simpler, mean-field model, which allows
for a more detailed analysis and more accurate simulations.

We consider the mean-field Hamiltonian given by
\begin{equation}
\label{eq_hamil_random}
H_{\epsilon} = - J  \vec{m} \cdot \vec{\sigma}
- \epsilon
 \vec{\eta}  \cdot {\vec{\sigma}},
\end{equation}
where, as $\vec{\eta}_i$ before, $$\vec{\eta} = \eta \cdot
\vec{e}_y
$$ is the quenched random field in the $y$-direction. Here $\eta$
is a scalar, symmetric random variable,
which we assume here to be Gaussian distributed with zero mean and
unit variance. $\epsilon$ \((>0)\) is the parameter (typically
small) that quantifies the strength of the randomness.

%
%
The corresponding mean field equation for magnetization is:
\begin{equation}
 {\vec{m}} = Av_\eta \left[
\frac{\int {\vec{\sigma}} \exp(\beta J {\vec{m}}\cdot {\vec{\sigma}} + \beta \epsilon \eta \sigma_y)  d{\vec{\sigma}}}
 {\int \exp(\beta J {\vec{m}}\cdot {\vec{\sigma}} + \beta \epsilon \eta \sigma_y) d{\vec{\sigma}}} \right].
\label{meanfield_mag_nonzeroepsi}
\end{equation}
Here $Av_\eta (\cdot)$ denotes the average over the disorder, i.e.
the integral over $\eta$ with the appropriate distribution (here
assumed to be unit normal).

\subsection{Numerical simulations}

\begin{figure}[tbp]
\begin{center}
\epsfig{figure= 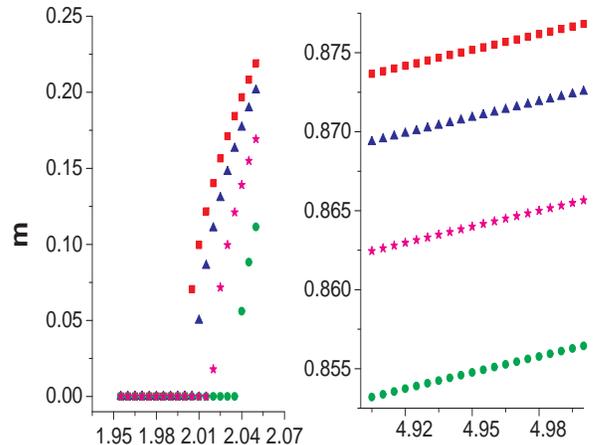,height=.3\textheight,width=0.47\textwidth}
\caption{(Color online.)
The magnetization  (\(m\)) with respect to $\beta J$ (horizontal axes): the (red) squares, (blue) triangles, (pink) stars, and
(green) circles are
for the system described by the Hamiltonian \(H_{\epsilon}\) with $\epsilon = 0$, $\epsilon = 0.1$, \(\epsilon =0.15\), and $\epsilon = 0.2$
respectively.
The figure on the left correspond to the behavior of the spontaneous magnetization near the critical temperature, while
that on the right correspond to the same at lower temperatures.
The decrease of magnitude of the magnetization due to disorder, is clearly of the order of $\epsilon^2$.}
\label{comparison_lowtemp}
\end{center}
\end{figure}

It follows from the symmetry of the distribution of $\eta$ that
all solutions of the equation (Eq. (\ref{meanfield_mag_nonzeroepsi})) have zero $Y$-component.  In the case of $\epsilon
\neq 0$, the system again does not magnetize at high temperature,
as in the case of $\epsilon = 0$. However,
there exists a critical temperature, below which a transverse
(with respect to the direction of the random field) magnetization
appears.
More precisely, there exists a \(\beta_c^{\epsilon}\) such that
for $\beta > \beta_c^{\epsilon}$, the magnetization equation has
two solutions with zero $Y$-components, whose $X$-components equal
$m$ and $-m$ (where $m > 0$)
(along with the trivial solution $m = 0$).
We will study the dependence of \(m\) on the temperature and on
the disorder strength \(\epsilon\).
Therefore, in contrast to the continuous set (a circle) of
solutions of the mean field equation in the system without
disorder (modelled by \(H_{XY}\)), in the disordered case there
are just two possible values of the spontaneous magnetization. In
Fig. \ref{comparison_hightemp}, the blue triangles represent the
magnetization for $\epsilon = 0.1$, while the red squares
correspond to \(\epsilon =0\). The same color code applies to Fig.
\ref{comparison_lowtemp}, where, in addition, pink stars represent
the case \(\epsilon =0.15\), and green circles---$\epsilon = 0.2$.
All the curves show two real solutions ($m$ and $- m$), of the
corresponding mean field equation, at low temperatures.
From  Figs. \ref{comparison_hightemp} and \ref{comparison_lowtemp}, it is clear that $m(\beta,\epsilon)$
is smaller than $m_0$ (see Eq. (\ref{dhorimaachh})), at low temperatures.
They coincide for high temperatures, as both of them are vanishing in that regime.
Numerical simulations show that the difference  $\delta(\beta,\epsilon) = m_0 - m(\beta,\epsilon)= m(\beta,0) - m(\beta,\epsilon)>0$
is of the order of $\epsilon^2$, in the regime of $\beta >
\beta_c^{\epsilon}$  (see Fig. \ref{comparison_lowtemp}).

\subsection{Scaling of critical temperature and magnetization with disorder: Perturbative approach}

In this subsection, we will study the mean field Hamiltonian in
Eq. (\ref{eq_hamil_random}), using perturbation theory and compare
the results with the numerical results of the last section.
Since, as argued earlier, the spontaneous magnetization can only
have nonzero $X$-component at any temperature, the mean field
equation (Eq. (\ref{meanfield_mag_nonzeroepsi})) reduces to
\begin{eqnarray}
 m = Av_\eta \left[
\frac{\int_{0}^{2 \pi} \cos \theta \exp(\beta J m \cos \theta + \beta \epsilon \eta \sin \theta)  d\theta}
 {\int_{0}^{2 \pi} \exp(\beta J m \cos \theta + \beta \epsilon \eta \sin \theta) d\theta} \right] \nonumber \\
\equiv F_\epsilon(m).
\label{meanfield_mag_nonzeroepsi_simplify}
\end{eqnarray}
Note that \(F_\epsilon (m) = F(m)\) for \(\epsilon =0\).

\subsubsection{Critical temperature}

To find the critical temperature, a similar method as in the case
of $\epsilon = 0$ (in the preceding section) is applied. The
condition for non-zero magnetization is now given by
\begin{equation}
F^{'}_{\epsilon}(0) = \frac{\beta J}{2} - \frac{\beta^3 \epsilon^2 J}{16} + o(\epsilon^2)>1,
\end{equation}
where \(o(\zeta^n)\) denotes a term which is of order higher than \(\zeta^n\).
This implies
\begin{equation}
\beta_c^\epsilon = \beta_c^0 - \frac{\epsilon^2}{2 J^3} + o(\epsilon^2).
\end{equation}
Therefore, we obtain negative $\epsilon^2$ corrections to the
critical temperature, as observed in the numerical simulations
(see Fig. \ref{comparison_lowtemp}).

\subsubsection{Scaling of magnetization near criticality}
\label{Amber-kella}

The magnetization goes to zero as the temperature approaches the
critical temperature. Let us now use perturbation techniques to
see the behavior of magnetization \(m\) near criticality.

Before considering the disordered case (\(\epsilon \ne 0\)), let
us first consider the case when $\epsilon = 0$.  In this case, we
expand $F(m)$ (r.h.s. of Eq. (\ref{meanfield_mag_simplify})) in
$m$ around $m = 0$:
\begin{equation}
\label{Taylor_nodis}
F(m) = m F^{'}(0) +  \frac{m^3}{6} F^{'''}(0) + \mbox{higher order terms}.
\end{equation}
Differentiating Eq. (\ref{meanfield_mag}) and considering the
resulting elementary integrals one can easily see that, by
symmetry, $F(0) = F^{''}(0) = 0$ and that
\begin{equation}
F^{'}(0) = \beta J/2, \quad
F^{'''}(0) = - 3\beta^3J^3/8.
\end{equation}
Putting these values in Eq. (\ref{Taylor_nodis}), we obtain that
the magnetization near $\beta_c^0$
equals
\begin{equation}
\label{Taylor_mag_nodis}
m_{0} = \frac{2\sqrt{2}}{J} \beta^{-\frac{3}{2}} (\beta - \beta_c^0)^{\frac{1}{2}}.
\end{equation}
plus higher order terms.

A similar technique can now be used to study the behavior of $m$
near $\beta_c$ in the presence of disorder. Again $F_\epsilon(0) =
F^{''}_\epsilon(0) = 0$.  Hence the expansion of $F_\epsilon(m)$
near zero will be
\begin{equation}
\label{Taylor_dis}
F_\epsilon(m) = mF^{'}_\epsilon (0)  +  \frac{m^3}{6} F^{'''}_\epsilon(0) + \mbox{higher order terms},
\end{equation}
where
\begin{eqnarray}
F^{'}_\epsilon(0)  &=& \left(\frac{\beta J}{2} - \frac{\beta^3 \epsilon^2 J}{16}\right), \nonumber\\
F^{'''}_\epsilon(0) & = & \beta^3 J^3 \left[\left(\frac{3}{8} - \frac{\beta^2 \epsilon^2}{32}\right) -
3\left(\frac{1}{2} - \frac{\beta^2 \epsilon^2}{16}\right)^2\right]\nonumber
\end{eqnarray}
Putting these derivatives in Eq. (\ref{Taylor_dis}), we obtain the correction of magnetization due to
disorder as
\[
m = m_{0} \left( 1 - \frac{1}{2 J^2 m_{0}^2} \epsilon^2 + o(\epsilon^2)\right).
\]
This is in full agreement with numerical simulations, that also showed a decrease of magnetization of order \(\epsilon^2\), in the
disordered case, as compared to the case when \(\epsilon=0\).

\subsubsection{A modified Bessel function and its expansion for large arguments}

In the following, we will have numerous occasions to use the
modified Bessel function
\begin{equation}
I_n(z) = \frac{1}{\pi}\int_{0}^{\pi} \exp(z \cos \vartheta) \cos(n \vartheta) d\vartheta,
\end{equation}
where $n$ is an integer, and $z$ is a real number, and the
expansion
of $I_n (z)$ for large $|z|$:  \cite{Abrohamawitz}
\begin{eqnarray}
\label{expansion_Bessel}
 I_n(z) &\approx& \frac{\exp(z)}{\sqrt{2 \pi z}}  \Big[ 1 - \frac{\mu -1}{8 z}
 +  \frac{(\mu -1)(\mu -9)}{2! (8 z)^2} \nonumber \\
&&- \frac{(\mu -1)(\mu - 9)(\mu -25)}{3! (8z)^3} + o \left(\left(1/z\right)^3\right)\Big], \nonumber \\
\end{eqnarray}
where \(n\) is fixed and $\mu = 4 n^2$. Actually, the function and
its expansion are true \cite{Abrohamawitz} for certain complex
ranges of the parameter \(z\). However, we will only use them for
real \(z\).  Here \(o\left(\left(1/\upsilon\right)^n\right)\)
denotes an expression, containing terms of order higher than $({1
\over n})^n$:   \(\left(1/\upsilon\right)^{n+1},
\left(1/\upsilon\right)^{n+2}, \ldots\).

\subsubsection{Scaling of magnetization at low temperature}

We now study behavior of $m$ at low temperatures, i.e. for large
$\beta$.

We again start from the case $\epsilon = 0$. Note that the
numerator and denominator of $F(m)$ (after some simple
modifications) are of the form of $I_n(z)$. Therefore, for large
$\beta$, we can use the asymptotics of the Bessel function in Eq.
(\ref{expansion_Bessel}) to obtain a low temperature expansion of
$F(m)$. Using this expansion, we obtain the following equation for
$m$, from Eq. (\ref{meanfield_mag_simplify}):
\begin{equation}
\label{eq_Bessel_m}
m^3 -m^2 + \frac{m}{2 \beta J}
+ o\left(1/\beta\right)=0.
\end{equation}
Since $m \rightarrow 1$ as $\beta \rightarrow \infty$, let us write $m$ as
\begin{equation}
 m = 1 - \frac{a_1}{\beta}
+ o\left(1/\beta\right).
\end{equation}
Putting this in Eq. (\ref{eq_Bessel_m}), we finally obtain the behavior of the magnetization for the case when \(\epsilon=0\),
 for large $\beta$:
\begin{equation}
 m_0 = 1 - \frac{1}{2 J \beta}
+ o\left(1/\beta\right).
\end{equation}

Using the same technique for the disordered case, we expand the
numerator and denominator of $F_\epsilon(m)$, using Eq.
(\ref{expansion_Bessel}), obtaining
\begin{eqnarray}
 m &=&1 - \frac{\epsilon^2}{2 J^2}-\frac{1}{\beta}
\left[\frac{1}{2 J} + \epsilon^2\left(\frac{3}{4J^3} - \frac{1}{2J^2}\right)\right] + o(1/\beta) \nonumber\\
&=& m_0 - \epsilon^2 \left\{ \frac{1}{2 J^2}+\frac{1}{\beta}
\left[  \left(\frac{3}{4J^3} - \frac{1}{2J^2}\right)\right]\right\} + o(1/\beta). \nonumber \\
\end{eqnarray}
As before, disorder leads to corrections of order \(\epsilon^2\)
to the magnetization at low temperature. This is again in
agreement with the numerical simulations (see Fig.
\ref{comparison_lowtemp}).

\section{Ferromagnetic XY model in a random field plus a constant field: Random field induced order}
\label{girgiti}

We have seen in the preceding section that a random field that
breaks the symmetry of the XY model, restricts possible
magnetization values to a discrete set. Although the system still
magnetizes, we no longer have continuously symmetry of the set of
solutions to the mean field equation, as a result of adding a
symmetry-breaking random field.
In this section we explore the effects of such a random field on a
system which already has a unique direction of the magnetization,
determined by a uniform magnetic field.

\begin{figure}[tbp]
\begin{center}
\epsfig{figure= 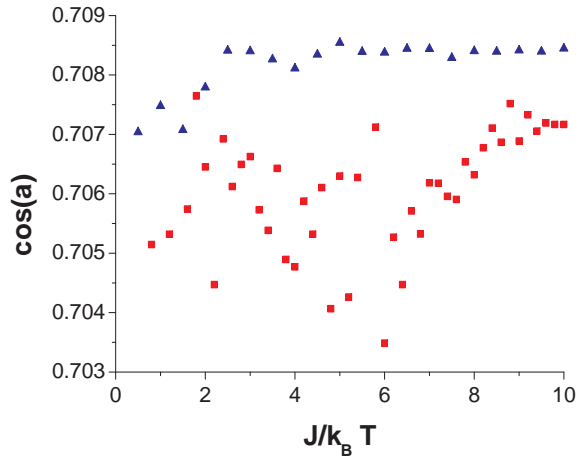,height=.3\textheight,width=0.47\textwidth}
\caption{(Color online.) Plot of \(\cos(a)\) with respect to \(\beta J\). Red squares represent the case when
the XY model has the  applied constant field \(\vec{h}\) with \(h=J\) and \(x= \pi/4\). Blue triangles are
for the same system but with the additional symmetry breaking random field of strength \(\epsilon = 0.1 J\).}
\label{cos_zetapiby4}
 \end{center}
\end{figure}

\begin{figure}[tbp]
\begin{center}
\epsfig{figure= 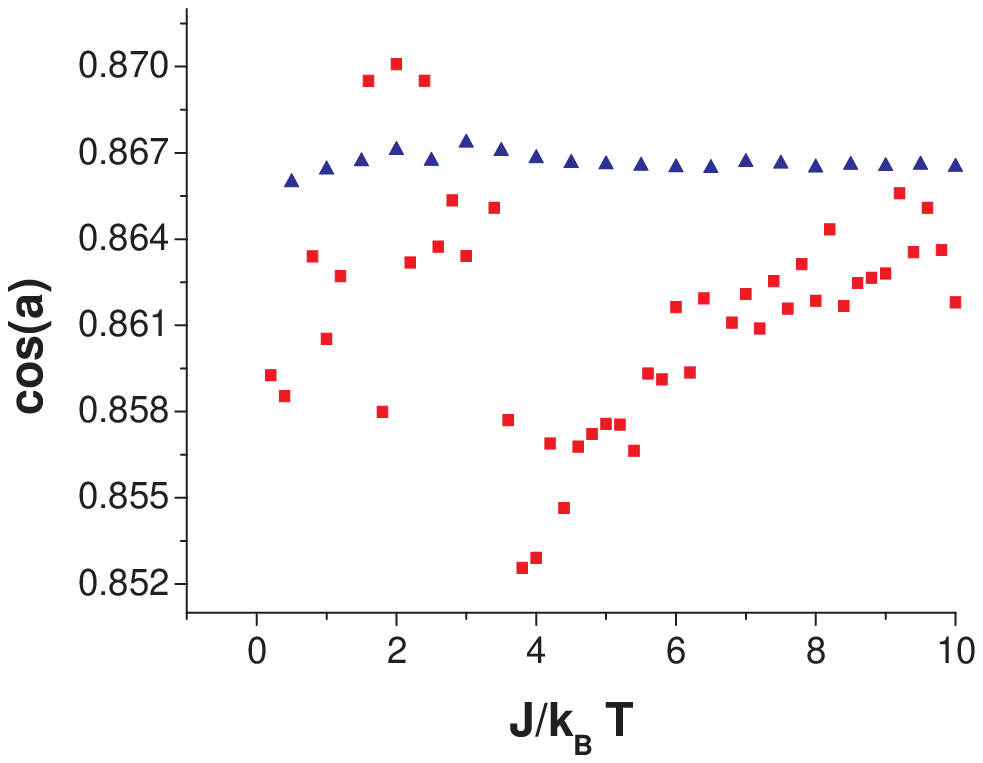,height=.3\textheight,width=0.47\textwidth}
\caption{(Color online.) This is the same plot as in Fig. \ref{cos_zetapiby4}, but for
\(x= \pi/6\).
}
\label{cos_zetapiby6}
 \end{center}
\end{figure}

\begin{figure}[tbp]
\begin{center}
\epsfig{figure= 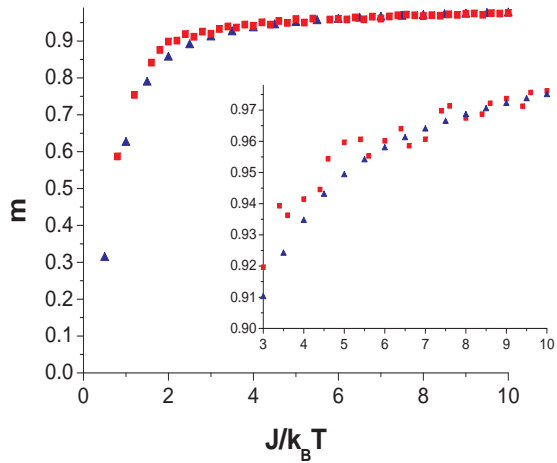,height=.3\textheight,width=0.47\textwidth}
\caption{(Color online.) Plot of the magnitude \(m\) of the magnetization with respect to \(\beta J\). Red squares represent the case when
the XY model has the  applied constant field \(\vec{h}\) with \(h=J\) and \(x= \pi/4\). Blue triangles are
for the same system but with the additional symmetry breaking random field of strength \(\epsilon = 0.1 J\).
}
\label{r_zetapiby4}
 \end{center}
\end{figure}

\begin{figure}[tbp]
\begin{center}
\epsfig{figure= 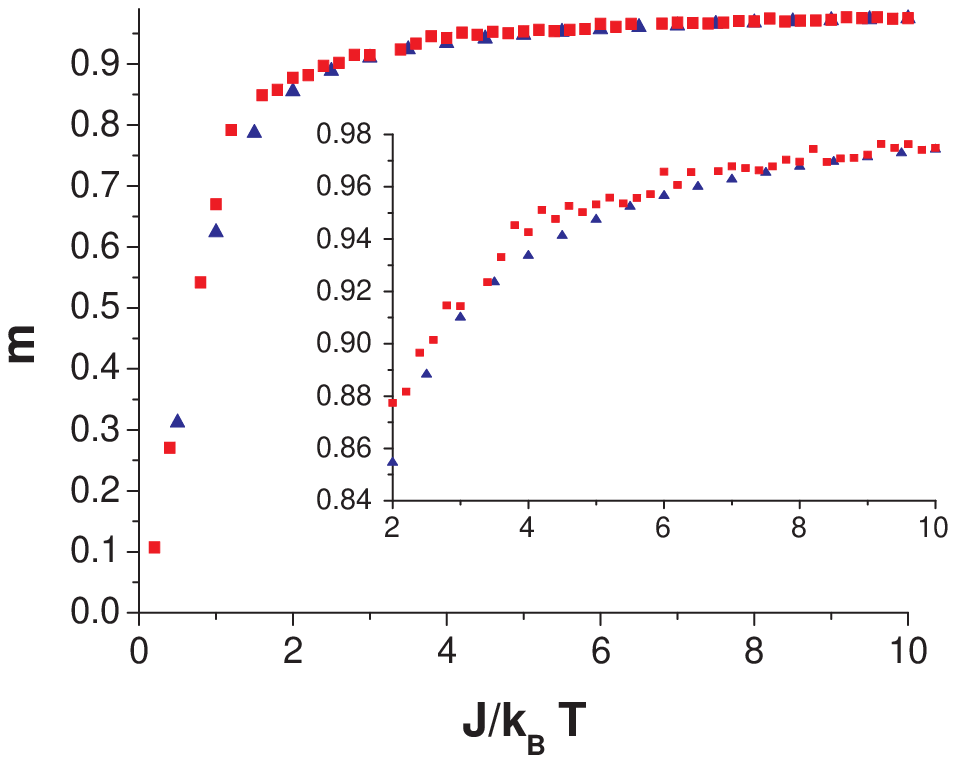,height=.3\textheight,width=0.47\textwidth}
\caption{(Color online.) This is the same plot as in Fig. \ref{r_zetapiby4}, but for
\(x= \pi/6\).
}
\label{r_zetapiby6}
 \end{center}
\end{figure}

First consider the case in which the planar symmetry in the XY
model is broken by applying a constant magnetic field
 \(\vec{h}\) alone.
That is, according to the general mean field strategy, we are
looking for the solutions of the following equation:
\begin{equation}
{\vec{m}} = \frac{\int {\vec{\sigma}} \exp(\beta J {\vec{m}}\cdot {\vec{\sigma}} + \beta {\vec{h}} \cdot  {\vec{\sigma}}) d{\vec{\sigma}}} {\int \exp(\beta J {\vec{m}}\cdot {\vec{\sigma}} + \beta {\vec{h}} \cdot {\vec{\sigma}}) d{\vec{\sigma}}}.
\label{mean_constfield}
\end{equation}
Let $ {\vec{h}} = (h\cos x, h\sin x)$. We suppose that \(0<h\leq 1\) and \(0\leq x \leq \pi/2\).
%
As expected, due to the applied constant field, the mean field equation has a unique solution of the magnetization $\vec{m}$ at all temperatures, and the
solution is a (positive) multiple of ${\vec{h}}$, but of reduced magnitude.
Red squares in  Figs. \ref{r_zetapiby4}, \ref{cos_zetapiby4}, \ref{r_zetapiby6}, and \ref{cos_zetapiby6},
correspond to the magnitude (\(m\)) and the cosine of the phase (\(a\)) of the magnetization \(\vec{m}\).

Let us now, in addition, apply a random field \(\{\eta\}\) in the
Y-direction.
The new mean field equation is
\begin{equation}
\label{1567}
{\vec{m}} = Av_{\eta}\left[\frac{\int {\vec{\sigma}} \exp(\beta J {\vec{m}}\cdot {\vec{\sigma}} + \beta {\vec{h}} \cdot  {\vec{\sigma}} +
\beta \epsilon \eta \sigma_y) d{\vec{\sigma}}} {\int \exp(\beta J {\vec{m}}\cdot {\vec{\sigma}} + \beta {\vec{h}} \cdot {\vec{\sigma}}
+ \beta \epsilon \eta \sigma_y) d{\vec{\sigma}}}\right].
\end{equation}
Here we have to solve the two simultaneous equations, given by Eq. (\ref{1567}),
to obtain the magnitude and the phase of the magnetization vector \(\vec{m}\).
In all the previous mean field (vector) equations, one could apriori predict the phase of the magnetization.
Just as in the case of a constant field \(\vec{h}\) and \(\epsilon =0\), again the solution remains unique.

Just as in the previous sections, we will now compare the
magnetization of the system without disorder (i.e. \(\epsilon=0\),
and for which the mean field equation is given by Eq.
(\ref{mean_constfield}))), with the system in which \(\epsilon \ne
0\) (and for which the mean field equation is given by
Eq.(\ref{1567}))keeping \(h\) strictly positive in both cases. Let
us denote the two Hamiltonians by \(H_h\) and \(H_{h,\epsilon}\)
respectively. We do the comparison by numerical simulations as
well as perturbatively at low temperatures (Sect.
\ref{Arga-ki-quila} below). A perturbation approach, similar to
the one in Sect. \ref{Amber-kella}, can be done at high
temperatures also. We refrain from doing it, as the high
temperature behavior in this case is not so interesting, in view
of absence of a phase transition.

\begin{figure}[tbp]
\begin{center}
\epsfig{figure= 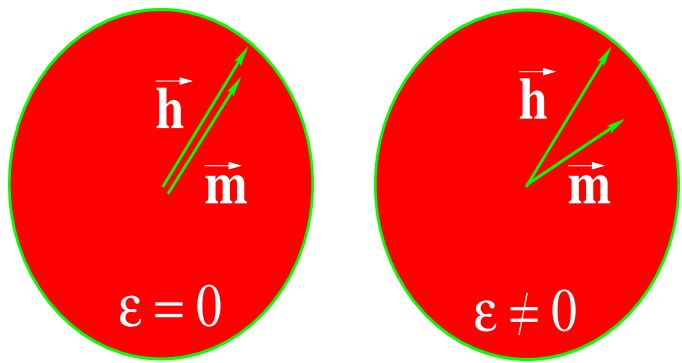,height=.2\textheight,width=0.5\textwidth}
\caption{(Color online.)
Schematic diagram of the magnetization of XY ferromagnets without and with disorder,
in the presence of a constant magnetic field.
The figure on the left indicates the behavior of \(\vec{m}\) in the presence of \(\vec{h}\), but when \(\epsilon =0\),
while the one on the right
is when there is a positive \(\epsilon\).
}
\label{schematic}
 \end{center}
\end{figure}

The length \(m\) of the magnetization vector is shrunk inthe
system described by \(H_{h,\epsilon}\), compared to the case in
which there is no disorder in the system (i.e. the one described
by \(H_h\)). This is seen from numerical simulations (see Figs.
\ref{r_zetapiby4} and \ref{r_zetapiby6}), as well as by
perturbation techniques at low temperatures.
In addition, numerical simulations (as shown in Figs.
\ref{cos_zetapiby4} and \ref{cos_zetapiby6}) show that the cosine
of the phase of the magnetization, i.e. $\cos(a)$ increases in
presence of the random field. Therefore, the phase of the
magnetization vector moves towards the X-direction (i.e. the
direction transverse to the applied random field). This is also
corroborated by perturbative approach at low temperatures. The
schematic diagram in Fig. \ref{schematic} shows the change of
behavior of the length and phase of the magnetization with and
without disorder, in the presence of a constant field.

The \(Y\)-component, \(m_y= m\sin a\), of the magnetization
has the same relative behavior as the length \(m\), in systems
described by \(H_h\) and \(H_{h,\epsilon}\), i.e. for small
$\epsilon > 0$ it is lower than for $\epsilon = 0$.

\begin{figure}[tbp]
\begin{center}
\epsfig{figure= 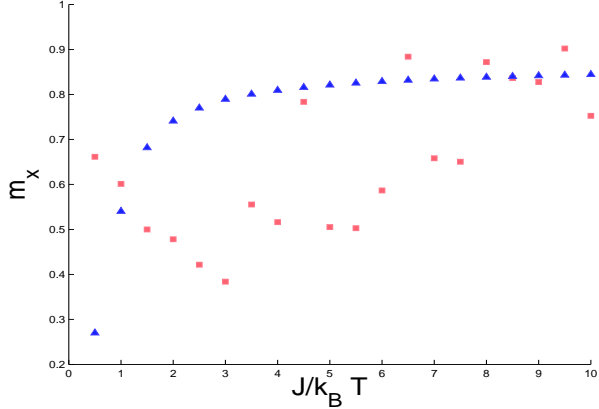,height=.25\textheight,width=0.45\textwidth}
\caption{(Color online.) 
Order from disorder: Red squares are the X-components of the magnetization in the absence of disorder,
while blue triangles are those in the presence of disorder $\epsilon = 0.1 J$. 
The constant field is present in both cases, with $x = \pi/6$. }
\label{mx}
 \end{center}
\end{figure}

However, the \(X\)-component, \(m_x=m\cos a\), of the magnetization, \(\vec{m}\), behaves in a very
interesting way.
Its value in the system described by \(H_{h,\epsilon}\) can be \textbf{both higher as well as lower} than its value in the system
described by \(H_h\).
The  numerical simulations in this case are given in Fig. \ref{mx}.

\subsection{Magnetization at low temperature: Perturbative approach}
\label{Arga-ki-quila}


To obtain the behavior of magnetization at low temperature, we will use the implicit function theorem, which we now state.
Let an equation \(f(x_1,x_2)=0\) of two variables \(x_1\) and \(x_2\) be such that
\(f(x_1,x_2)=0\) at \((x_1,x_2)=(x_1^0,x_2^0)\). \(x_2\) is in general an unknown function of \(x_1\).
But we may still understand the character of \(\frac{dx_2}{dx_1}|_{x_1=x_1^0}\), by
using the fact that (under certain regularity conditions on \(f\) near \((x_1^0,x_2^0)\))
\begin{equation}
\frac{\partial f}{\partial x_1}\Big|_{(x_1^0,x_2^0)} + \frac{\partial f}{\partial x_2}\Big|_{(x_1^0,x_2^0)}
\frac{dx_2}{d x_1}\Big|_{(x_1^0,x_2^0)}=0.
\end{equation}
The usual statement of the implicit function theorem is that when
${\partial f \over \partial x_2}$ is nonzero at
\((x_1^0,x_2^0)\)), we can solve the equation $f(x_1, x_2) = 0$
for $x_2$ uniquely near this point and the derivative of the
resulting function ($x_2$ as a function of $x_1$) at $x_1^0$ can
then be calculated from the above equation.  However, in the case
when the first derivatives vanish at a certain point, we can use a
simple extension of it to calculate the second derivatives. Such a
situation appears in the calculations below of the second
derivatives of the magnetization with respect to \(\epsilon\).

The mean field equations that we work with here can be written in
the form
\begin{equation}
\vec{m}=\frac{1}{\beta J} \nabla_{\vec{m}} \Gamma,
\end{equation}
where
\begin{equation}
\nabla_{\vec{m}} \equiv \left( \frac{\partial }{\partial m_x}, \frac{\partial }{\partial m_y} \right),
\end{equation}
with
\begin{equation}
\Gamma = \log_e \int \exp\left(-\beta H_h \right) \quad \mbox{or}
\quad \log_e \int \exp\left(-\beta H_{h,\epsilon} \right).
\end{equation}

It follows from symmetry of the distribution of $\eta$ that
$\vec{m}$ is an even function of $\epsilon$ and, consequently
\(\frac{dm_x}{d\epsilon}\) and \(\frac{dm_y}{d\epsilon}\) vanish
at \(\epsilon =0\).

It follows that
\begin{eqnarray}
\label{6787}
\frac{d^2m_x}{d\epsilon^2} \left[ 1 - \frac{1}{\beta J} \frac{\partial^2 \Gamma}{\partial m_x^2} \right] &=&
                 \frac{1}{\beta J} \left[ \frac{\partial^3 \Gamma}{\partial^2 \epsilon\partial m_x} +
                                          \frac{\partial^2 \Gamma}{\partial m_y \partial m_x} \frac{d^2m_y}{d\epsilon^2} \right] \nonumber \\
\frac{d^2m_y}{d\epsilon^2} \left[ 1 - \frac{1}{\beta J} \frac{\partial^2 \Gamma}{\partial m_y^2} \right] &=&
                 \frac{1}{\beta J} \left[ \frac{\partial^3 \Gamma}{\partial^2 \epsilon\partial m_y} +
                                          \frac{\partial^2 \Gamma}{\partial m_y \partial m_x} \frac{d^2m_x}{d\epsilon^2} \right], \nonumber \\
\end{eqnarray}
where all the total and partial derivatives are taken at
\(\epsilon=0\).  The above system of equations can be solved for
the second (total) derivatives \(\frac{d^2m_x}{d\epsilon^2}\) and
\(\frac{d^2m_y}{d\epsilon^2}\), at \(\epsilon =0\), once we can
find the partial derivatives at \(\epsilon=0\).

\begin{figure}[tbp]
\begin{center}
\epsfig{figure= 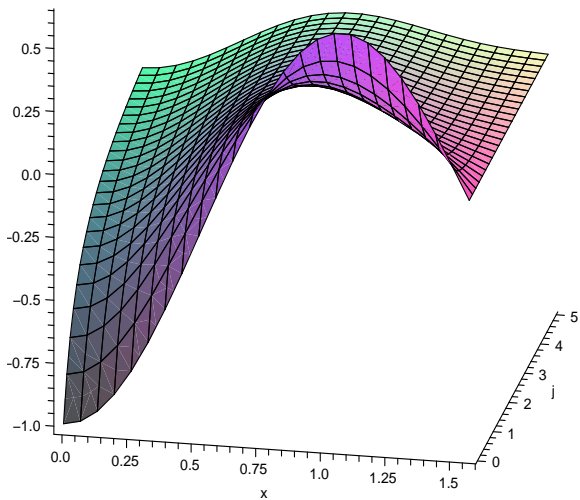,height=.4\textheight,width=0.5\textwidth}
\caption{(Color online.)
Plot of the function \(X(x,j)\)  with respect to \(x\) and   \(j=J/h\).
Note that there are ranges of the \((x,j)\), for which the function \(X\) is positive.
This fact gives rise to the phenomenon of random field induced order in the system described by the Hamiltonian \(H_{h,\epsilon}\).
}
\label{fig_X}
 \end{center}
\end{figure}

\begin{figure}[tbp]
\begin{center}
\epsfig{figure= 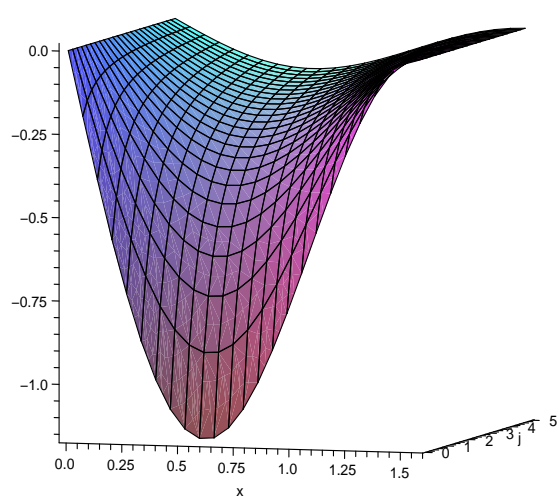,height=.4\textheight,width=0.5\textwidth}
\caption{(Color online.)
Plot of the function \(Y(x,j)\)  with respect to \(x\) and   \(j\). Clearly, it is negative for the entire range of \(x\) and \(j\).
}
\label{fig_Y}
 \end{center}
\end{figure}

The partial derivatives in Eq. (\ref{6787}) are calculated using
the following strategy. We have
\begin{equation}
\frac{1}{\beta J} \frac{\partial \Gamma}{\partial m_x} = Av_\eta \langle \cos \theta \rangle,
\end{equation}
where for any observable \(A\), \(\langle A \rangle \) is the
Gibbs average
\begin{equation}
\langle A \rangle = \frac{\int A \exp (-\beta H)}{\int \exp(-\beta
H)},
\end{equation}
with \(H\) being the relevant Hamiltonian (\(H_h\) or
\(H_{h,\epsilon}\)).  Of course, in the case of the system
described by the Hamiltonian \(H_h\), the quenched averaging with
respect to \(\eta\) is not required.  Using this notation we have,
differentiating the formula for $\Gamma$ twice,
\begin{equation}
\frac{1}{\beta J} \frac{\partial}{\partial \epsilon}\frac{\partial \Gamma}{\partial m_x}
= Av_\eta \left[ \beta \eta \left( \langle \cos \theta \sin \theta\rangle - \langle \cos \theta\rangle \langle \sin \theta\rangle\right)\right],
\end{equation}
and
\begin{eqnarray}
&&\frac{1}{\beta J} \frac{\partial^2}{\partial \epsilon^2}\frac{\partial \Gamma}{\partial m_x}
= Av_\eta \Big[ \beta^2 \eta^2 \Big( \langle \cos \theta \sin^2 \theta\rangle - \nonumber \\
&& 2\langle \cos \theta \sin \theta\rangle \langle \sin \theta\rangle + 2\langle \cos \theta \rangle \langle \sin \theta\rangle^2
 - \langle \cos \theta \rangle \langle \sin^2 \theta\rangle\Big)\Big]. \nonumber \\
\end{eqnarray}
We expand these partial derivatives with respect to \(1/\beta\),
at \(\epsilon=0\), using the expansion of the modified Bessel
function \(I_n\).  After some calculations, we obtain
\begin{equation}
\frac{d^2m_x}{d\epsilon^2}\Big|_{\epsilon=0}= \frac{1}{h^2}X\left(x,\frac{J}{h}\right) + \mbox{order of } \frac{1}{\beta},
\end{equation} and
\begin{equation}
\frac{d^2m_y}{d\epsilon^2}\Big|_{\epsilon=0}= \frac{1}{h^2}Y\left(x,\frac{J}{h}\right) + \mbox{order of } \frac{1}{\beta},
\end{equation}
where the functions 
\(X\) and \(Y\) are given by (for \(j=J/h\))
\begin{eqnarray}
X(x,j)=\frac{A D + BC }{DE -C^2}  \label{bordaX}\\
Y(x,j)=\frac{BD+AC}{DE - C^2}, \label{bordaY}
\end{eqnarray}
where 
\begin{widetext}
\begin{eqnarray}
A(x,j)= \frac{1}{(j+1)^3} \Big[ -\frac{1}{32} (  (3j+1)\cos x + (45j+63)\cos 3x  ) - \frac{5}{8} (  j+3  ) \sin 2x \sin x
+ \frac{1}{4} (  j+2  ) \cos x \cos 2x  \nonumber \\
-\frac{1}{8} (  3j-7  ) \cos x \sin^2 x  \Big],  \\
B(x,j)= \frac{1}{(j+1)^3} \Big[ -\frac{1}{32} (  3(3j+1)\sin x + (45j+63)\sin 3x  ) + \frac{3}{2} (  j+2  ) \cos 2x \sin x
+ \frac{3}{8} (  j+3  ) \sin^3 x   \Big], 
\end{eqnarray}
\end{widetext}
\begin{eqnarray}
C(x,j)= - \frac{j \cos x \sin x}{ j+1} \\
D(x,j)= \frac{j \cos^2 x +1}{j+1} \\
E(x,j)= \frac{j \sin^2 x +1}{j+1}
\end{eqnarray}
We plot the functions \(X\) and \(Y\) in Figs. \ref{fig_X} and \ref{fig_Y}.



Therefore, at low temperatures, we have, up to order
\(\epsilon^2\):
\begin{eqnarray}
m_x= m_x|_{\epsilon=0} + \epsilon^2\left(\frac{1}{h^2}X\left(x,\frac{J}{h}\right) + \mbox{order of } \frac{1}{\beta}\right), \\
m_y= m_y|_{\epsilon=0} + \epsilon^2\left(\frac{1}{h^2}Y\left(x,\frac{J}{h}\right) + \mbox{order of } \frac{1}{\beta}\right).
\end{eqnarray}
From Fig. \ref{fig_Y}, it is clear that the \(Y\)-component of the magnetization always decreases in
the presence of disorder.
However, Fig. \ref{fig_X} shows that there are ranges in the parameter space \((x,j)\),
for which the
quenched averaged \(X\)-component, \(m_x\), of the magnetization \emph{increases} in the presence of
 disorder, compared to the case when there is
no disorder. As noted before, this is in agreement with our
numerical simulations.

We have also considered the effect of disorder on the length \(m\)
and phase \(a\) of the magnetization.  For the phase, we consider
the expansion of \(\tan(a) = \frac{m_y}{m_x}\), which is given by
\begin{equation}
\tan (a)= \frac{m_y}{m_x}\Big|_{\epsilon=0} + \epsilon^2 \frac{d^2}{d\epsilon^2}\left(\frac{m_y}{m_x}\right)\Big|_{\epsilon=0} + o(\epsilon^2),
\end{equation}
with
\begin{eqnarray}
\frac{d^2}{d\epsilon^2}\left(\frac{m_y}{m_x}\right)\Big|_{\epsilon=0} &=&
\frac{m_x \frac{d^2m_y}{d\epsilon^2} - m_y \frac{d^2m_x}{d\epsilon^2}}{m_x^2}\Big|_{\epsilon=0}
 \nonumber \\
               &=& \frac{1}{m_x^2 |_{\epsilon=0}} \frac{1}{h^2} S\left( x, j\right) + \mbox{order of } \frac{1}{\beta}, \nonumber \\
\end{eqnarray}
where 
\begin{equation}
S(x,j)= Y(x,j) \cos x - X(x,j) \sin x,
\end{equation}
with \(X\) and \(Y\) being given by Eqs. (\ref{bordaX}) and (\ref{bordaY}).

\begin{figure}[tbp]
\begin{center}
\epsfig{figure= 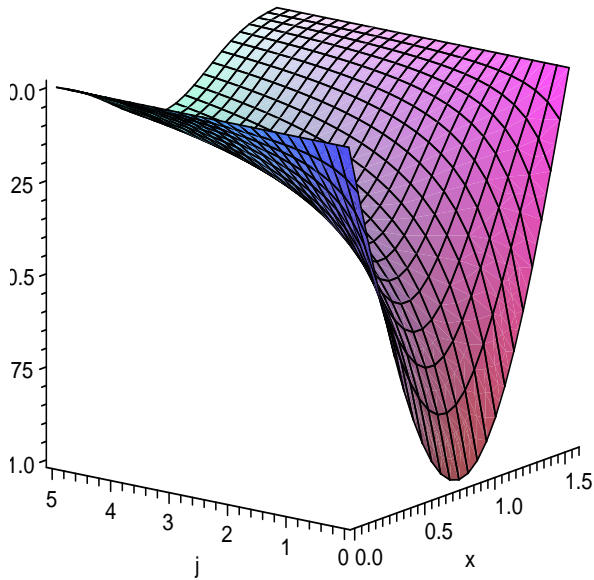,height=.4\textheight,width=0.5\textwidth}
\caption{(Color online.)
Plot of the function \(S(x,j)\)  with respect to \(x\) and   \(j\). It is again negative for the entire range of \(x\) and \(j\).
This is in agreement with the numerical results in Figs. \ref{cos_zetapiby4} and \ref{cos_zetapiby6}.
}
\label{fig_sign_3d}
 \end{center}
\end{figure}

\begin{figure}[tbp]
\begin{center}
\epsfig{figure= 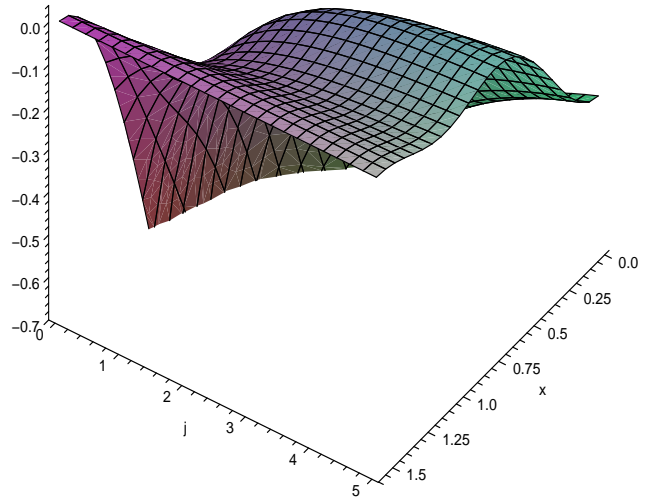,height=.4\textheight,width=0.5\textwidth}
\caption{(Color online.)
Plot of the function \((Xm_x + Ym_y)|_{\epsilon=0}\) with respect to \(x\) and \(j\).
It is negative for all \(x\) and \(j\),
and numerical results in Figs. \ref{r_zetapiby4} and \ref{r_zetapiby6} corroborate this effect.
}
\label{fig_length}
 \end{center}
\end{figure}

As shown in Fig. \ref{fig_sign_3d},  \(S\left( x, j\right)\) is
negative for all \(x\) and \(j\). Consequently, the phase \(a\)
always bends towards the \(X\)-direction in the presence of
disorder (since \(\tan(a)\) decreases in $a$, \(\cos(a)\)
increases), as we have already seen in simulations.  Note that
\(0\leq a\leq \pi/2\).
The square of the length of the magnetization is given by (up  to order \(\epsilon^2\))
\begin{eqnarray}
&&m_x^2 + m_y^2 = (m_x^2 + m_y^2)|_{\epsilon=0} \nonumber \\
&& + 2 \epsilon^2 \left((Xm_x + Ym_y)|_{\epsilon=0} + \mbox{order of } \frac{1}{\beta}\right).
\end{eqnarray}
As seen on Fig. \ref{fig_length}, \((Xm_x + Ym_y)|_{\epsilon=0}\)
is always negative, showing that the length of the magnetization
decreases in the presence of disorder.
Note that the behavior of the length and phase obtained
perturbatively, matches what is shown schematically in Fig.
\ref{schematic}.



\section{Discussion}
\label{alochona}

To summarize, we have considered classical systems of two
dimensional spins, and studied the interplay between continuous
symmetry and symmetry-breaking quenched disordered field, in the
mean field approximation.  We found that in case of a system in a
uniform magnetic field, disorder may enhance one component of the
order parameter.


In this paper, we have explicitly considered only the situation
when the spins are two-dimensional. It is natural to ask analogous
questions for three-dimensional spins with continuous symmetry.
Below we argue that for the canonical system of this kind---the
classical Heisenberg model---the behavior is similar to that of
the XY model.


To  study the behavior of magnetization in the lattice Heisenberg
model in the presence of disorder, we put at all sites random
fields in the Y-direction. The Hamiltonian of the resulting
disordered system is given by


\begin{equation}
 H_{H} = - J \sum_{|i -j| = 1} \vec{\sigma}_i \cdot \vec{\sigma}_j - \epsilon \sum_i \vec{\eta} \cdot \vec{\sigma}_i,
\end{equation}
where $\vec{\eta} = \eta \cdot \vec{e}_y $. Here $\vec{\sigma}$
are now $3$D unit vectors.  The mean field Hamiltonian is again
\begin{equation}
 \vec{m} =  - J\vec{m} \cdot \vec{\sigma} - \epsilon \vec{\eta} \cdot \vec{\sigma},
\end{equation}
 where  $\vec{m}$ is the magnetization, and this time we parametrize
\(\vec{\sigma}\) as $(\sin \theta \cos \phi, \cos\theta \sin\phi,
\cos\theta)$.  Therefore, the mean field equation reads
\begin{equation}
 \vec{m} = Av_\eta \left[
\frac{\int {\vec{\sigma}} \exp(\beta J {\vec{m}}\cdot {\vec{\sigma}} + \beta \epsilon \eta \sigma_y)  \sin\theta d\theta d\phi}
 {\int \exp(\beta J {\vec{m}}\cdot {\vec{\sigma}} + \beta \epsilon \eta \sigma_y) \sin\theta d\theta d\phi} \right] \equiv F_H(m),
\end{equation}
where we have chosen
\[
\vec{m} = (m \sin \psi \cos \chi, m \sin \psi \sin \chi, m \cos \psi).
\]

Consider first the case, when $\epsilon = 0$. By symmetry, the solutions
of the mean field equation in this case form a sphere for $\beta > \beta_c^{0H}$.


Suppose that the radius of the sphere is \(m_0^H\). To find
$\beta_c^{0H}$ analytically, we can use an argument that is
similar to the one that we have used
in the case of the XY model. The existence of critical temperature gives the condition
\begin{equation}
 F^{'}_H (0) = \frac{\beta J}{3} > 1
\end{equation}
which implies
\[\beta_c^{0H} = \frac{3}{J}.\]
The behavior of the magnetization as  \(\beta\) approaches $\beta_c^{0H}$, is given by
\begin{equation}
 m_0^H = \frac{\sqrt{5}}{\sqrt{2}} \frac{1}{J} (\beta_c^{0H})^{-\frac{3}{2}} (\beta - \beta_c^{0H})^{\frac{1}{2}}.
\end{equation}
Note that in the case of the XY model, we also found a similar behavior of the magnetization near its critical temperature.

In the presence of disorder, i.e. \(\epsilon \ne 0\), we obtain the correction to the critical temperature as
\begin{eqnarray}
\beta_c^{\epsilon H} &=& \frac{3}{J} + \frac{9}{5} \frac{\epsilon^2}{J^3} + o(\epsilon^2), \nonumber\\
&=& \beta_c^{0H} + \frac{9}{5}\frac{\epsilon^2}{J^3} + o(\epsilon^2),
\end{eqnarray}
which is again qualitatively siumilar to the situation in the XY model.
The magnetization near the critical
temperature is also decreased by an order of $\epsilon^2$. The relation between the magnetization  without disorder ($m_0^H$),
and that  in the presence of disorder ($m^H$) is
\begin{equation}
 m^H = m_0^H  \left(1 - \frac{1}{12 J^2 (m_0^H)^2} \epsilon^2\right).
\end{equation}

\vspace{0.1cm}

\begin{acknowledgments}
We acknowledge support from the DFG
(SFB 407, SPP1078 and SPP1116, 436POL),
Spanish
Ministerio de Ciencia y Tecnolog{\' i}a
grants FIS-2005-04627 and ``Ram{\'o}n y Cajal'',
Acciones Integradas and
Consolider QOIT,
the ESF Program QUDEDIS, and EU IP SCALA. J. W. was partially
supported by the NSF grant DMS 0623941. 	
\end{acknowledgments}


\end{document}